\documentclass[fp,twocolumn]{jpsj3}
\usepackage{amssymb}
\usepackage{txfonts}

\inst{$^1$Key Laboratory for Microstructural Material Physics of
Hebei Province, School of Science, Yanshan University, Qinhuangdao 066004, China \\
$^2$Department of Physics, Xinzhou Teachers University, Xinzhou 034000, China\\
$^3$Department of Physics, Tangshan Normal University, Tangshan
063000, China\\$^4$Department of Physics and Astronomy, University
of Pittsburgh, Pittsburgh,Pennsylvania 15260, USA}
 \abst{We
investigate the ground-state structures of dipolar
spin-orbit-coupled Bose-Einstein condensates in a toroidal trap.
Combined effects of dipole-dipole interaction (DDI) and spin-orbit
coupling (SOC) on the ground states of the system are discussed. A
ground-state phase diagram is obtained as a function of the SOC and
DDI strengths. As two new degrees of freedom, the DDI and SOC can be
used to obtain the desired ground-state phases and to control the
phase transition between various ground states. In particular, the
system displays exotic topological structures and spin textures,
such as half-quantum vortex, vortex string, vortex necklace, complex
vortex lattice including giant vortex and hidden antivortex chains,
different skyrmions, meron (half-skyrmion)-antimeron
(half-antiskyrmion) necklace, and composite meron-antimeron
lattice.}

\begin{document}

\title{Ground states of dipolar spin-orbit-coupled Bose-Einstein condensates
in a toroidal trap}
\author{Hui Yang$^{1}$$^{,}$$^{2}$, Qingbo Wang$^{1}$$^{,}$$^{3}$, and
Linghua Wen$^{1}$$^{,}$$^{4}$\thanks{%
linghuawen@ysu.edu.cn}}
\maketitle

\section{Introduction}

The past few decades have witnessed unprecedented advances in experimental
and theoretical investigations on ultracold atomic gases \cite{Chin,Stamper}%
. In general, the contact interatomic interactions that can be controlled by
Feshbach resonance play a key role in the physical properties of
Bose-Einstein condensates (BECs). For most of alkali atomic BECs, the
two-body contact interaction between atoms is crucially determined by the $s$%
-wave scattering length and the other interactions can be neglected.
However, for the BECs made of atoms with large magnetic dipole moments there
are not only contact interaction but also strong dipole-dipole interaction
(DDI) \cite{Lahaye1}. The magnetic DDI can be attractive or repulsive
depending on the orientation of atomic dipole moments, the relative position
of particles and the geometric structure of the system. Recent experimental
and theoretical studies of dipolar quantum gases with chromium \cite{Lahaye2}%
, dysprosium \cite{Lu}, and erbium atoms \cite{Aikawa} show that DDI can
have pronounced effects on the mean-field ground states, Bogoliubov spectra,
many-body stationary states, and dynamics of the BECs or degenerate Fermi
gases \cite{Santos,Yi,Kadau,Kato,Cha,Borgh,Zou,Wenzel,Jia}. These remarkable
effects include ferrosuperfluid \cite{Lahaye2}, droplet crystals \cite{Kadau}%
, dipolar superfluid \cite{Wenzel}, flat Landau levels \cite{Zhou}, and
roton mode in a dipolar quantum gas \cite{Chomaz}.

On the other hand, spin-orbit coupled quantum gases have also become one of
the focuses in cold atom physics in recent years \cite{Zhai}. Compared with
the natural solid state system with unavoidable impurities and disorder, the
ultracold atomic gases with spin-orbit coupling (SOC) provide a new testing
ground for studying novel SOC physics and potential applications due to
their ultrahigh purity, experimental precise controllability and excellent
theoretical description \cite{Lin,Qu,ZWu,LHuang,Kolkowitz,Li}, particularly
for BECs \cite%
{Hu,YZhang,Zhu,Ruokokoski,Aftalion,YXu,Jiang,Poon,Sakaguchi,SGautam}. In a
realistic physical system, ultracold atomic gases are confined in an
external potential. As a matter of fact, the presence of various external
potentials leads to rich physics because different trapping potentials can
dramatically affect the physical properties of the BECs. For instance, the
breathing mode and scissors mode in a harmonic trap \cite{Pethick}, the
quantum phase transition from a superfluid to a Mott insulator in an optical
lattice \cite{Greiner}, and the hidden vortices \cite{Wen1,Mithun,Price,Wen2}
and macroscopic quantum self-trapping \cite{Smerzi} in a double-well
potential have been revealed in the literature. Recently, a new type of
nonsimply connected trapping potential, i.e., a toroidal potential, has
become experimentally available due to the development of different
techniques. The toroidal potential allows nontrivial topological structures,
unusual quantum dynamics and tunable features \cite%
{Eckel,Wood,XFZhang,White,Wang,Helm}.

In this work, we study the ground-state properties of dipolar BECs with SOC
in a toroidal trap. To obtain the ground state of the system, we numerically
calculate the coupled Gross-Pitaevskii (GP) equations by using standard
imaginary-time propagation method \cite{YZhang,Wen3}. The results show that
one can obtain the desired ground-state phases by controlling the DDI
strength and SOC strength and regulating the phase transition between
different ground states. In addition, we find that the system supports rich
topological structures and exotic spin textures which are expected to be
observed in the future cold-atom experiments. In the rest of this paper, it
is organized as follows. The theoretical model is presented in Sec. 2. The
results and discussion are shown in Sec. 3. Finally we give a brief summary
in Sec. 4.

\section{Model}

We consider quasi-two-dimensional (quasi-2D) two-component dipolar BECs with
Rashba SOC in a toroidal trap and with strong confinement in the $z$%
-direction. In the frame of mean-field theory, the dynamics of the system is
described by the coupled 2D GP equations,%
\begin{eqnarray}
i\hbar \frac{\partial \psi _{j}}{\partial t} &=&[-\frac{\hbar ^{2}}{2m}%
\nabla ^{2}+V(r)+g_{j}\left\vert \psi _{j}\right\vert
^{2}+g_{j(3-j)}\left\vert \psi _{(3-j)}\right\vert ^{2}  \nonumber \\
&&+v_{so}+\phi _{j}+\phi _{j(3-j)}]\psi _{j},  \label{1-1}
\end{eqnarray}%
where $\psi _{j}(r)$ $(j=1,2)$ denote the two-component wave functions, and
the normalization condition is given by $\int [\left\vert \psi
_{1}\right\vert ^{2}+\left\vert \psi _{2}\right\vert ^{2}]dr=N$ with $N$
being the number of atoms. $g_{j}$ $=\sqrt{8\pi }\hbar ^{2}a_{j}/ma_{z}$ $%
(j=1,2)$ and $g_{12}$ $=g_{21}=\sqrt{8\pi }\hbar ^{2}a_{12}/ma_{z}$
represent the quasi-2D intra- and intercomponent coupling strengths \cite%
{Kasamatsu}, where we have assumed that the two component atoms have the
same mass $m$, $a_{z}=\sqrt{\hbar /m\omega _{z}}$with $\omega _{z}$ being
the harmonic trap frequency in the $z$ direction, and $a_{j}$ $(j=1,2)$ and $%
a_{12}$ denote the $s$-wave scattering lengths between intra- and
intercomponent atoms, respectively. The Rashba SOC is expressed by $%
v_{so}=-i\hbar k(\hat{\sigma}_{x}\partial _{y}-\hat{\sigma}_{y}\partial
_{x}) $ with $\hat{\sigma}_{x,y}$ being Pauli matrices and $k$ being the
isotropic SOC strength. The toroidal trap is given by \cite{Wang,Cozzini}%
\begin{equation}
V(r)=\frac{1}{2}\hbar \omega _{\bot }\left[ V_{0}\left( \frac{r^{2}}{%
a_{0}^{2}}-r_{0}\right) ^{2}\right] ,  \label{potential}
\end{equation}%
where $\omega _{\bot }$ is the radial oscillation frequency, $r=\sqrt{%
x^{2}+y^{2}}$, and $a_{0}=\sqrt{\hbar /m\omega _{\bot }}$. $V_{0}$ and $%
r_{0} $ represent the central height and the width of the toroidal trap,
respectively. $\phi _{j}(x,y)$ $(j=1,2)$ and $\phi _{12}(x,y)=\phi
_{21}(x,y) $ denote the DDIs of intraspecies and interspecies, respectively,
and they can be expressed as \cite{Lahaye1,SYi1}%
\begin{eqnarray}
\phi _{j} &=&c_{j}\int d\mathbf{r}^{\prime }U_{dd}(\mathbf{r}-\mathbf{r}%
^{\prime })\left\vert \psi _{j}(\mathbf{r}^{\prime })\right\vert ^{2},
\label{IntraDDI} \\
\phi _{j(3-j)} &=&c_{12}\int d\mathbf{r}^{\prime }U_{dd}(\mathbf{r}-\mathbf{r%
}^{\prime })\left\vert \psi _{(3-j)}(\mathbf{r}^{\prime })\right\vert ^{2}.
\label{DDI}
\end{eqnarray}%
Here $c_{j}=\mu _{0}\mu _{j}^{2}/4\pi $ $(j=1,2)$ and $c_{12}=\mu _{0}\mu
_{1}\mu _{2}/4\pi $ are magnetic dipole constants of intraspecies and
interspecies, $\mu _{0}$ is the vacuum magnetic permeability, and $\mu _{j}$
$(j=1,2)$ represent magnetic dipole moments of the two components,
respectively. Let us consider the simplest case in which the dipoles are
polarized. Then $U_{dd}(\mathbf{R})$ is given by \cite{MUeda}%
\begin{equation}
U_{dd}(\mathbf{R})=(1-3\cos ^{2}\theta )/R^{3},  \label{ddi}
\end{equation}%
where $\theta $ is the angle between the direction of polarization and the
relative position of atoms.

For the sake of discussion, we assume DDI exists only in component 1 and the
magnetic dipole moment is in the $z$ direction, i.e., $\phi _{2}=\phi
_{12}=\phi _{21}=0$. In view of the numerical computations, it is convenient
to introduce the following notations $\tilde{r}=r/a_{0},\tilde{t}=\omega
_{\bot }t,$ $\tilde{V}(r)=V(r)/\hbar \omega _{\bot },$ $\tilde{\phi}%
_{1}=\phi _{1}/\hbar \omega _{\bot }$, $\tilde{\psi}_{j}=\psi _{j}a_{0}/%
\sqrt{N}(j=1,2)$, $\beta _{jj}=g_{j}Nm/\hbar ^{2}$ $(j=1,2)$ and $\beta
_{12}=g_{12}Nm/\hbar ^{2}$, and then we obtain the dimensionless coupled GP
equations,%
\begin{eqnarray}
i\partial _{t}\psi _{1} &=&\left( -\frac{1}{2}\nabla ^{2}+V+\beta
_{11}\left\vert \psi _{1}\right\vert ^{2}+\beta _{12}\left\vert \psi
_{2}\right\vert ^{2}+\phi _{1}\right) \psi _{1}  \nonumber \\
&&+k\left( \partial _{x}-i\partial _{y}\right) \psi _{2},  \label{com1} \\
i\partial _{t}\psi _{2} &=&\left( -\frac{1}{2}\nabla ^{2}+V+\beta
_{22}\left\vert \psi _{2}\right\vert ^{2}+\beta _{12}\left\vert \psi
_{1}\right\vert ^{2}\right) \psi _{2}  \nonumber \\
&&-k(\partial _{x}+i\partial _{y})\psi _{1}.  \label{com2}
\end{eqnarray}%
For simplicity the tilde is omitted throughout this paper. Here $\beta _{jj}$
$(j=1,2)$ and $\beta _{12}$ are dimensionless intra- and interspecies
coupling strengths for the contact interactions, and $k$ is the
dimensionless SOC strength as mentioned before. For the present dipolar
BECs, the system interactions include \textit{s}-wave interactions, DDI and
SOC. It is convenient to introduce a dimensionless quantity to describe the
magnitude of the DDI relative to the contact interaction \cite%
{MUeda,Kawaguchia,XFZhang2},
\begin{equation}
\varepsilon _{dd}=a_{dd}/a_{1}=\frac{\mu _{0}\mu _{1}^{2}m}{12\pi \hbar
^{2}a_{1}},  \label{Interaction Ratio}
\end{equation}%
where $a_{dd}\ $denotes the scattering length characterizing the DDI. Here
we assume that the dipoles are polarized along the $z$ direction, i.e.,
these dipoles are arranged side by side along the $z$ axis. In this context,
the DDI becomes an isotropic repulsion (or attraction) and can be equivalent
to a contact interaction. Therefore, the DDI can be rewritten as the form of
effective contact interaction \cite{Kawaguchia,XFZhang2}%
\begin{equation}
\phi _{1}=\beta _{11}\varepsilon _{dd}\left\vert \psi _{1}\right\vert ^{2},
\label{Effective DDI}
\end{equation}%
where the DDI is repulsive for $\varepsilon _{dd}>0$ and attractive for $%
\varepsilon _{dd}<0$ when $a_{1}>0$. Obviously the total interaction
coefficient between atoms for component $1$ can be expressed by $%
(1+\varepsilon _{dd})\beta _{11}$. Thus one can obtain and control different
ground-state phases by varying the DDI strength $\varepsilon _{dd}$, the SOC
strength $k$, and the interaction strengths $\beta _{11}$, $\beta _{22}$ and
$\beta _{12}$.

In order to further understand the topological properties of the system, we
adopt a nonlinear Sigma model \cite{Kasamatsu2,Wang} in which a normalized
complex-valued spinor $\mathbf{\chi }=[\chi _{1},\chi _{2}]^{T}$ with $|\chi
_{1}|^{2}+|\chi _{2}|^{2}=1$ is introduced. The total density of the system
is expressed by $\rho =|\psi _{1}|^{2}+|\psi _{2}|^{2}$, where the
corresponding two-component wave functions are $\psi _{1}=\sqrt{\rho }\chi
_{1}$ and $\psi _{2}=\sqrt{\rho }\chi _{2}$, respectively. The spin density
is given by $\mathbf{S}=\overline{\mathbf{\chi }}\mathbf{\sigma \chi }$ with
$\mathbf{\sigma }=(\sigma _{x},\sigma _{y},\sigma _{z})$ being the Pauli
matrices. The components of $\mathbf{S}$ are defined as%
\begin{eqnarray}
S_{x} &=&\chi _{1}^{\ast }\chi _{2}+\chi _{2}^{\ast }\chi _{1},  \label{Sx}
\\
S_{y} &=&i(\chi _{2}^{\ast }\chi _{1}-\chi _{1}^{\ast }\chi _{2}),
\label{Sy} \\
S_{z} &=&|\chi _{1}|^{2}-|\chi _{2}|^{2},  \label{Sz}
\end{eqnarray}%
with $|\mathbf{S}|=\sqrt{S_{x}^{2}+S_{y}^{2}+S_{z}^{2}}=1$. The spacial
distribution of the topological structure of the system can be characterized
by the topological charge density%
\begin{equation}
q(r)=\frac{1}{4\pi }\mathbf{S\bullet }\left( \frac{\partial \mathbf{S}}{%
\partial x}\times \frac{\partial \mathbf{S}}{\partial y}\right) ,
\label{TopologicalChargeDensity}
\end{equation}%
and the topological charge is given by $Q=\int q(r)dxdy.$

\section{Results and discussion}

In the absence of DDI and SOC, the two-component BECs in a toroidal trap
support two typical ground-state phases: phase mixing (component mixing) and
phase separation (component demixing) \cite{White,Wang}. The two different
types of ground states can be achieved by controlling the intra- and
interspecies contact interactions. Here we focus on the case of strong
confinement of the toroidal trap, where $V_{0}=5$ and $r_{0}=3$. Using the
standard imaginary time evolution \cite{YZhang,Wen3} based on the
Peaceman--Rachford algorithm \cite{Wen4,Peaceman}, we can obtain the
ground-state structures of two-component BECs with DDI and SOC in a toroidal
trap by numerical computation. The main idea of the Peaceman--Rachford
algorithm is to convert a 2D problem into two 1D problems, and the algorithm
can be easily extended to 3D cases. The imaginary-time propagation method
with the Peaceman--Rachford algorithm has good convergence, high accuracy
and strong stability. A test for the method is given by the fast convergent
values of the total energy and component wave functions of the system. A
second test of convergence and accuracy is provided by the virial theorem,
which fixes rigorous relationships among the different contributions to the
kinetic and potential energies of the system. In addition, it can also be
tested by the same convergent results for different trial wave functions.
All the tests have been confirmed in our numerical simulation. In order to
highlight the new effects resulted from the DDI and SOC, without loss of
generality we fix the interaction parameters $\beta _{11}=100$, $\beta
_{22}=150$ and $\beta _{12}=200$ throughout this paper, and vary the DDI
strength $\varepsilon _{dd}$ or the SOC strength $k$. In fact, our
simulation shows that there exist similar phase structures for the other
combination cases of contact interaction parameters.

\subsection{Ground-state structures}

Inspired by Ref. \cite{Kato}, we give the ground-state phase diagram spanned
by the SOC strength $k$ and the DDI strength $\varepsilon _{dd}$ in Fig. 1.
There are eight different phases marked by A-H, which differ in terms of
their density profiles and phase distributions. In the following discussion,
we will give a detailed description of each phase. The density and the phase
profiles of the eight different phases A-H in Fig. 1 are shown in Figs.
2(a)-2(e) and Figs. 3(b), 3(c) and 3(f), respectively.

\begin{figure}[tbp]
\centerline{\includegraphics*[width=10cm]{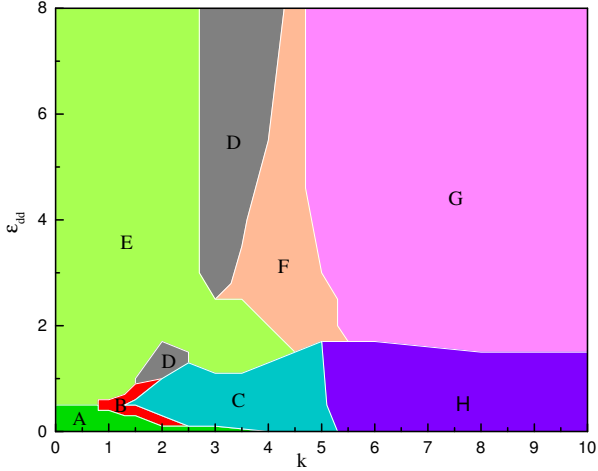}}
\caption{(Color online) Ground-state phase diagram of the spin-orbit coupled
dipolar BEC with respect to $k$ and $\protect\varepsilon _{dd}$ for $\protect%
\beta _{11}=100$, $\protect\beta _{22}=150$ and $\protect\beta _{12}=200$.
There are eight different phases marked by A-H.}
\label{Figure1}
\end{figure}

We start from the case where the SOC and the DDI are weak, which is denoted
by the region A in Fig. 1. In this phase, the density distributions of the
two components display layer-segregated symmetry preserving annular
structures [see the top two rows in Fig. 2(a)], which is mainly due to the
strong repulsive interspecies contact interaction $\beta _{12}^{2}>\beta
_{11}\beta _{22}$ and the tight binding of the toroidal trap. From the
corresponding phase distributions in rows 3 and 4 of Fig. 2, we observe that
the value of the phase changes continuously from $-\pi $ (blue) to $\pi $
(red), and the end point of the boundary between a $\pi $ phase line and a $%
-\pi $ phase line corresponds to a phase defect (i.e., a vortex with
anticlockwise rotation or an antivortex with clockwise rotation). The
topological structure of the system in Fig. 2(a) is a half-quantum vortex
where there is no phase defect in the dipolar component (component 1) while
there is an obvious vortex in the non-dipolar component (component 2). Here
the half-quantum vortex is different from the conventional
Anderson--Toulouse coreless vortex \cite{Anderson,Matthews} due to the
presence of the toroidal trap. For the latter case, the vortex core of one
component is usually filled with the other nonrotating component. With the
slightly strong inclusion of SOC and DDI, the A phase transforms to the B
phase, as shown in Fig. 1. The component densities evolve into separated
stripe-like patterns or block-like patterns, and some phase defects are
generated in the system [see Fig. 2(b)]. With an increase in the strength of
the SOC, the C phase emerges as the ground state, as shown in Fig. 1.
Typical density distributions are shown in Fig. 2 (c), in which the
interlaced and separated petal structures in the component density
distributions become quite pronounced, in the mean time, there is a triply
quantized giant vortex in the central region of the phase distribution of
the dipolar component, followed successively by a hidden antivortex necklace
\cite{Wen1,Wen2} in the external region and ghost vortex necklace as well as
ghost antivortex necklace \cite{Kasamatsu} in the further boundary regions.
By comparison, the central region of the phase profile of component 2 is
occupied by a triply quantized giant antivortex with a hidden vortex
necklace distributing in the outer region and ghost antivortex as well as
ghost vortex necklaces in the further boundary regions. These phase defects
constitute nucleated vortices and antivortices on the whole. This feature is
evidently different from the case of two-component spin-orbit coupled BECs
in a harmonic trap \cite{Aftalion}. For the latter case, only conventional
stripe phase and segregated symmetry preserving phase with giant skyrmion
can be created when $\beta _{12}^{2}>\beta _{11}\beta _{22}$. Physically,
here the special petal phase is a result of the combined effect of DDI, SOC,
toroidal trap and strong interspecies repulsion.

\begin{figure}[tbp]
\centerline{\includegraphics*[width=8.5cm]{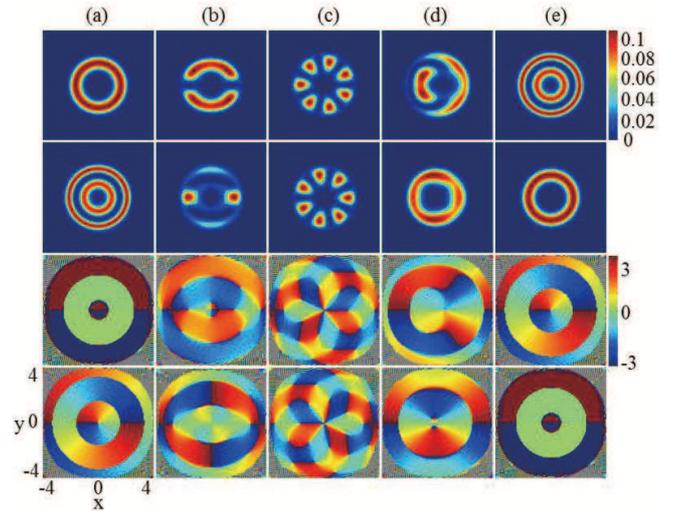}}
\caption{(Color online) Ground states of two-component BECs with DDI and SOC
in a toroidal trap, where $\protect\beta _{11}=100$, $\protect\beta %
_{22}=150 $, $\protect\beta _{12}=200$, and $k=2$. (a) $\protect\varepsilon %
_{dd}=0$, (b) $\protect\varepsilon _{dd}=0.2$, (c) $\protect\varepsilon %
_{dd}=0.5$. (d) $\protect\varepsilon _{dd}=1.5$ and (e) $\protect\varepsilon %
_{dd}=2$. The rows from top to bottom denote $\left\vert \protect\psi %
_{1}\right\vert ^{2}$, $\left\vert \protect\psi _{2}\right\vert ^{2}$, arg$%
\protect\psi _{1}$, and arg$\protect\psi _{2}$, respectively. The unit
length is $a_{0}$. The value of the phase in the phase distributions (bottom
two rows) changes continuously from $-\protect\pi $ (blue) to $\protect\pi $
(red), and the end point of the boundary between a $\protect\pi $ phase line
and a $-\protect\pi $ phase line corresponds to a phase defect (i.e., a
vortex with anticlockwise rotation or an antivortex with clockwise
rotation). }
\end{figure}

\begin{figure*}[tbh]
\centerline{\includegraphics*[width=10cm]{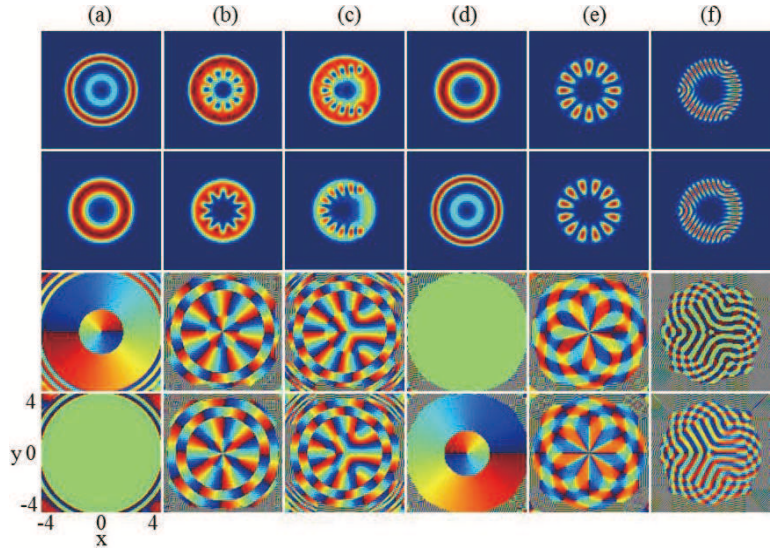}}
\caption{(Color online) Ground states of two-component BECs with various DDI
and SOC strengths in a toroidal trap, where the contact interaction
parameters are the same as those in Fig. 1. (a) $\protect\varepsilon _{dd}=6$%
, $k=0.5$, (b) $\protect\varepsilon _{dd}=6$, $k=4.5$, (c) $\protect%
\varepsilon _{dd}=6$, $k=5$, (d) $\protect\varepsilon _{dd}=0.3$, $k=0.5$,
(e) $\protect\varepsilon _{dd}=0.3$, $k=4$, and (f) $\protect\varepsilon %
_{dd}=0.3$, $k=10$. The rows from top to bottom denote $\left\vert \protect%
\psi _{1}\right\vert ^{2}$, $\left\vert \protect\psi _{2}\right\vert ^{2}$,
arg$\protect\psi _{1}$, and arg$\protect\psi _{2}$, respectively. The unit
length is $a_{0}$. The value of the phase in the phase distributions (bottom
two rows) changes continuously from $-\protect\pi $ (blue) to $\protect\pi $
(red), and the end point of the boundary between a $\protect\pi $ phase line
and a $-\protect\pi $ phase line corresponds to a phase defect (i.e., a
vortex with anticlockwise rotation or an antivortex with clockwise
rotation). }
\label{Figure3}
\end{figure*}

In the limit of weak SOC and relatively strong DDI, the C phase transforms
to the D phase, as shown in Fig. 1. The density and phase distributions are
shown in Fig. 2 (d) where the petals in the two component densities become
integrated and most of the phase defects disappear. In particular, two
vortex dipoles (vortex-antivortex pairs) instead of a giant vortex (or
antivortex) are formed in the density hole of component 2, which is
indicated in the phase distribution of component 2. Although the density
distributions in the B phase and the C phase do not have rotational
symmetry, they still maintain good axial symmetry concerning the $y=0$ axis
(or both the $y=0$ axis and the $x=0$ axis). In the presence of large
repulsive DDI the ground states of component 1 and component 2 in this
system are exchanged approximately with those in phase A, which is indicated
by the E phase in Fig.1. Typical density and phase distributions of such a
phase are shown in Fig. 2 (e). In addition, we find that different
combinations of parameters $\beta _{11}$ and $\varepsilon _{dd}$ with the
fixed other parameters may lead to the same ground-state structures in view
of the analytical effective interaction $(1+\varepsilon _{dd})\beta _{11}$
in the dipolar component. For instance, the ground state of the system in
the case of $\beta _{11}=120$ and $\varepsilon _{dd}=0.25$ is the same with
that in the case of $\beta _{11}=100$ and $\varepsilon _{dd}=0.5$, where the
ground-state structure for the former case is not shown here for the sake of
conciseness and avoiding repetition. Furthermore, our simulation shows that
the ground-state structure of the system in the presence of attractive DDI
above the critical value for collapse is similar to that in Fig. 2(a).

To get a deeper physical insight into this system, we study the ground-state
phase diagram with strong repulsive DDI ($\varepsilon _{dd}\geq 2.5$). With
the increase of the SOC, the phase transforms from the E phase to the D
phase, as shown in Fig. 1. For the relatively strong SOC ($k\geq 2.7$), the
F phase emerges as the ground state of the system, as shown by the pink
region in Fig. 1. Typical density and phase distributions of such a phase
are shown in Fig. 3(b), where there is a giant vortex (a multiquantized
vortex) \cite{Fetter} in the trap center surrounded by an antivortex
necklace in the outer region in each component. As the SOC is further
increased, the F phase transforms to the G phase (see the light purple
region of Fig. 1). In the G phase, interlaced and crescent-shaped visible
antivortex strings are formed in both the component densities, where there
is a large density hole in the central region of each component [see Fig.
3(c)]. The large density hole is not a usual giant vortex or antivortex but
a hidden vortex pair consisting of two singly quantized hidden vortices \cite%
{Wen1,Mithun,Price,Wen2} as shown in the phase distributions of Fig. 3(c).
This phase exists for strong SOC and strong DDI, and occupies the largest
region of the ground-state phase diagram in Fig. 1. Essentially, the
irregular structures of the vortices and antivortices in Fig. 3(c) are
caused by the complicated interplay among the DDI, SOC, contact interactions
and the toroidal confinement.

Finally, we move to another limit of weak DDI. When the SOC increases, the
phase transforms from A phase to B phase and then to C phase, and finally to
H phase as shown in Fig. 1. In the case of $\varepsilon _{dd}=0.3$ and $k=4$%
, each component density develops into a necklace-like structure which is
composed of 12 petals along the azimuthal direction [see Fig. 3(e)]. The
similar necklace structure has also been found in the recent literature \cite%
{White}, where only odd petals are observed. By comparison, not only the
necklace structure with odd number of petals but also that with even number
of petals can be observed easily in the present system due to the presence
of DDI [see Fig. 2(c) and Fig. 3(e)]. In addition, from the phase
distribution of component 2, one can see that there is no any phase defect
in the density hole. However, there exists a hidden vortex-antivortex
necklace in the ring area where the petals are located, followed by several
complicated ghost vortex necklaces or ghost vortex-antivortex necklaces in
the outer region of the atom cloud. In particular, a giant vortex (a five
quantized vortex) is distributed in the central region of component 1, and a
hidden vortex necklace as well as a hidden vortex-antivortex necklace
consisting of vortex-antivortex pairs are formed in the annular petal
region. At the same time, the external region of the cloud is occupied by
several composite ghost vortex-antivortex necklaces. \cite{White}. Under the
limit of strong SOC, the H phase emerges as the ground state, as shown in
the Fig. 1. The density and phase distributions of H phase are shown in Fig.
3(f), where some petals are connected into curved stripes and the density
distributions are evolved into several domains. Almost all the phase defects
via the form of vortex-antivortex pairs are distributed randomly in the
outer boundary region of the cloud except for a singly quantized vortex in
the trap center for component 2.

The z component of the orbital angular momentum, $\left\langle
L_{z}\right\rangle =\int \mathbf{\Psi }^{\dagger }(\mathbf{r})(xp_{y}-yp_{x})%
\mathbf{\Psi }(\mathbf{r})d\mathbf{r}$ with $\mathbf{\Psi }(\mathbf{r}%
)=(\psi _{1}$,$\psi _{2})^{\text{T}}$, is important in considering this
system. Figure 4 displays $\left\langle L_{z}\right\rangle $ as a function
of $k$ for $\varepsilon _{dd}=1$ being fixed. The E phase has less angular
momentum due to there being only one vortex in component 2. The first rapid
increase in $\left\langle L_{z}\right\rangle $ occurs in the D phase $%
(1.5\leq k\leq 2)$. The angular momentum of the C phase has increasing
trend, because the two components have increasing multiplequantum vortices
which carry angular momentum. In the H phase, $\left\langle
L_{z}\right\rangle $ decrease at $k$ $\simeq 5.5$ as the central region of
component 2 has no any phase defect. The maximum of $\left\langle
L_{z}\right\rangle $ is attained at $k$ $\simeq 6.5$, because the two
component have many vortices which carry a lot of angular momentum. When $k$
$\simeq 7.5$, $\left\langle L_{z}\right\rangle $ decreases and the reason is
similar to the case of $k$ $\simeq 5.5$.

\begin{figure}[tbp]
\centerline{\includegraphics*[width=7.5cm]{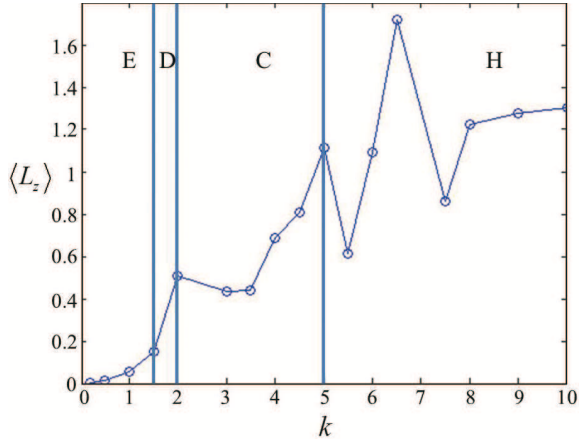}}
\caption{(Color online) Orbital angular momentum $\left\langle
L_{z}\right\rangle $ as a function of k for $\protect\varepsilon _{dd}=1$.
The vertical lines separate the phases and the solid curve indicates the
trend.}
\label{Figure4}
\end{figure}

From the above analysis and discussion, the ground-state structure of the
system can be transformed by adjusting the DDI strength $\varepsilon _{dd}$
or the SOC strength $k$ when the other parameters are fixed, which indicates
that the DDI and SOC can be used as two new degrees of freedom to achieve
the desired ground-state phases and to control the phase transition between
different ground states.\ In addition, the exotic topological structures
[especially shown in Figs. 2(b)-2(d) and Figs. 3(b), 3(c), 3(e) and 3(f)]
are quite different from the ground-state structures observed in
conventional SOC BECs \cite{Zhai,Lin,Hu,YZhang,Zhu,Ruokokoski,White},
dipolar BECs \cite{Yi,Wenzel}, and rotating two-component BECs with or
without SOC (DDI) \cite{Aftalion,Pethick,Wen1,Kasamatsu,Fetter}.

\subsection{Spin textures}

\begin{figure}[tbp]
\centerline{\includegraphics*[width=8cm]{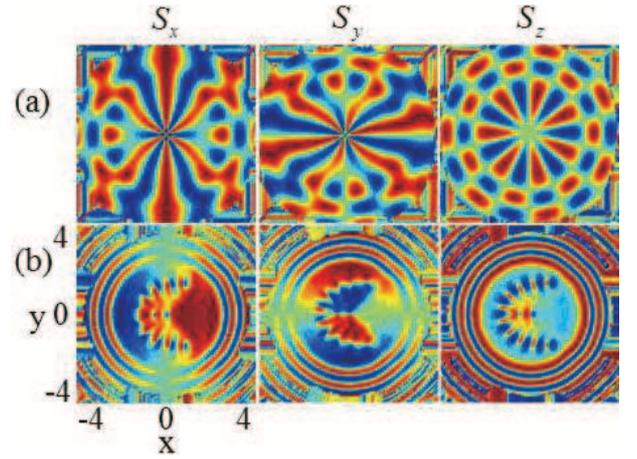}}
\caption{(Color online) Spin densities of dipolar spin-orbit coupled BECs in
a toroidal trap. (a) $\protect\varepsilon _{dd}=0.5$, $k=2$, and (b) $%
\protect\varepsilon _{dd}=6$, $k=5$. The columns (from left to right) denote
$S_{x},S_{y}$ and $S_{z}$ components of the spin density vector,
respectively. The corresponding ground states in (a) and (b) are shown in
Fig. 2(c) and Fig. 3(c), respectively. The unit length is $a_{0}$. }
\end{figure}

The typical spin density distributions are shown in Fig. 5, where $\beta
_{11}=100$, $\beta _{22}=150$ and $\beta _{12}=200$. The DDI and SOC
strengths in Figs. 5(a) and 5(b) are $\varepsilon _{dd}=0.5$, $k=2$, and $%
\varepsilon _{dd}=6$, $k=5$, respectively. The corresponding density and
phase distributions are given in Fig. 2(c) and Fig. 3(c), respectively. In
the spin representation, the blue region in the spin component $S_{z}$
represents spin-down ($S_{z}=-1$) and the red region denotes spin-up ($%
S_{z}=1$). For the case of weak DDI [see Fig. 5(a)], spin component $S_{x}$
shows an even parity distribution along both the $x$ direction and the $y$
direction while the situation is the reverse for $S_{y}$, i.e., $S_{y}$
displays an odd parity distribution along both the two directions. However, $%
S_{z}$ obeys an even parity distribution along the $x$ direction and an odd
parity distribution along the $y$ direction. By comparison, for the case of
strong DDI [see Fig. 5(b)], $S_{x}$ and $S_{z}$ display the even parity
distribution along the $x$ direction while $S_{y}$ shows the odd parity
distribution along the $x$ direction, where the even or odd parity
distributions along the $y$ direction are lost because of the long-range and
anisotropic feature of the DDI. Furthermore, one can notice that there is an
obvious petal-like pattern along the radial direction in the central region
and two distinct necklace-like patterns with increasing radius in the outer
region in the spin component $S_{z}$ of Fig. 5(a). The alternating
appearance of the blue and red petals along the radial direction as well as
that of the blocks along both the azimuthal direction and the radial
direction means that some regular spin domains are formed in the spin
representation. The boundary region between a blue petal (or block) and a
red petal (or block) actually constitutes a spin domain wall with $%
\left\vert S_{z}\right\vert \neq 1$. It is well known that for a system of
nonrotating two-component BECs the domain wall is a typically classical N%
\'{e}el Wall in which the spin flips only along the vertical direction of
the wall. However, our numerical simulation shows that the spin in the
region of domain wall flips not only along the vertical direction (azimuthal
direction) of the domain wall but also along the domain-wall direction
(radial direction), which indicates that here the observed domain wall is a
new type of domain wall and allows to be realized under current experimental
technical conditions.

\begin{figure*}[tbh]
\centerline{\includegraphics*[width=13cm]{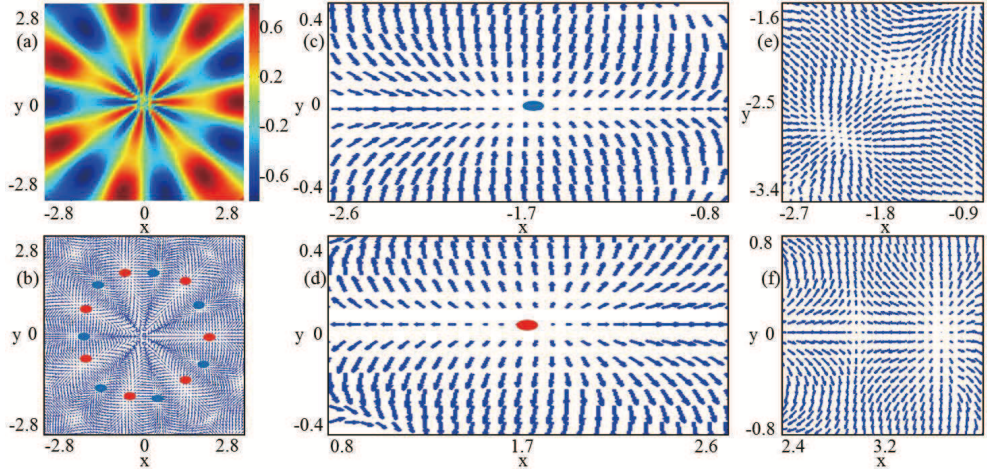}}
\caption{(Color online) Topological charge density and spin texture, where $%
\protect\varepsilon _{dd}=0.5$ and $k=2$. (a) topological charge density,
(b) the corresponding spin texture, and (c)-(f) the local amplifications of
the spin texture. Here the blue spots denote the half-antiskyrmions
(antimerons), and red spots represent the half-skyrmions (merons). The unit
length is $a_{0}$. }
\end{figure*}

Displayed in Figs. 6(a) and 6(b) are the topological charge density and the
spin texture of the system, where $\varepsilon _{dd}=0.5$, $k=2$, and the
contact interaction parameters are the same as those in Figs. 1-5. Figs.
6(c)-6(f) are the typical local enlargements of the spin texture in Fig.
6(b). The corresponding ground state and three spin components are given in
Fig. 2(c) and Fig. 5(a), respectively. In Figs. 6(b)-6(d), the blue spots
denote radial-in half-antiskyrmions (antimerons) \cite{Kasamatsu2,Shi} with
local topological charge $Q=-0.5$, and the red spots represent radial-out
half-skyrmions (merons) \cite{Skyrme,Mermin} with local topological charge $%
Q=0.5$. These meron-antimeron\ pairs (half-skyrmion and half-antiskyrmion
pairs) constitute an alternating complex and fascinating meron
(half-skyrmion)-antimeron (half-antiskyrmion) necklace, which, to the best
of our knowledge, has not yet been reported elsewhere. In addition, there
are some skyrmions (or antiskyrmions) \cite{Shi,Skyrme} with local
topological charge $\left\vert Q\right\vert =1$ distributed outside of the
meron (half-skyrmion)-antimeron (half-antiskyrmion) necklace, and typical
local amplifications are given in Figs. 6(e) and 6(f). As a matter of fact,
Fig. 6(e) describes a hyperbolic-radial(in) antiskyrmion with topological
charge $Q=-1$ while Fig. 6(f) denotes a hyperbolic-radial(out) skyrmion with
$Q=1$, where the former case can be regarded as a combination of two
antimerons (half-antiskyrmions) and the latter case can be considered as a
combination of two merons (half-skyrmions). Physically, the skyrmions or
merons (half-skyrmions) in the spin textures of the present system are
associated with the vortex structures in the component density profiles, and
the particle density should obey the continuity condition due to the quantum
fluid nature of the BECs.

\begin{figure}[tbp]
\centerline{\includegraphics*[width=8cm]{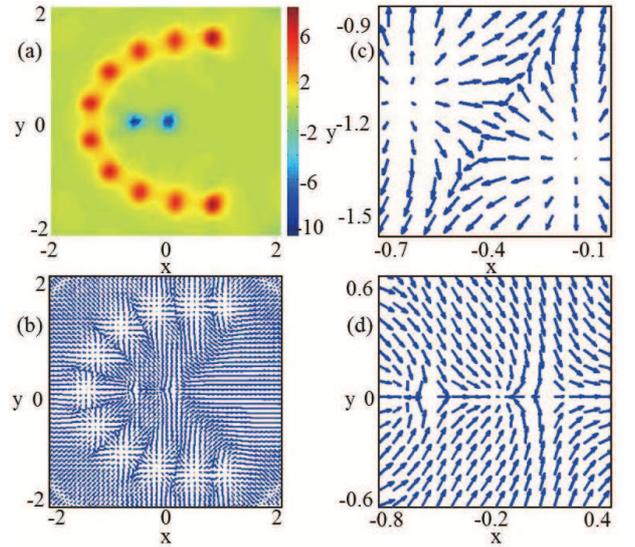}}
\caption{(Color online) Topological charge density and spin texture, where $%
\protect\beta _{11}=100$, $\protect\beta _{22}=150$, $\protect\beta %
_{12}=200 $, $\protect\varepsilon _{dd}=6$, and $k=5$. (a) topological
charge density, (b) the corresponding spin texture, and (c)-(d) the local
amplifications of the spin texture. The unit length is $a_{0}$. }
\end{figure}

Figure 7(a) shows the topological charge density in the case of $\beta
_{11}=100$, $\beta _{22}=150$, $\beta _{12}=200$, $\varepsilon _{dd}=6$, and
$k=5$, where the ground state and three spin components are displayed in
Fig. 3(c) and Fig. 5(b), respectively. The corresponding spin texture is
shown in Fig. 7(b), and the typical local amplifications are given in Figs.
7(c) and 7(d). Our numerical calculation demonstrates that the two local
enlargements in Fig. 7(c) are two merons (half-skyrmions) with each local
topological charge being $Q=0.5$, which indicates that the arc configuration
of the spin texture in Fig. 7(b) is a curved meron (half-skyrmion) chain. In
the mean time, the computation result shows that the two local topological
charges in Fig. 7(d) are both $Q=-0.5$, which means that the two spin
defects in Fig. 7(b) [see also Fig. 7(a)] along the $y=0$ axis in the area
surrounded by the arc configuration constitute an obvious antimeron pair (a
half-antiskyrmion pair) made of two antimerons (half-antiskyrmions). Thus
the curved meron chain and the antimeron pair jointly form a particular
composite meron-antimeron lattice, which has not been observed in previous
literature. The interesting and exotic spin textures as mentioned in Figs. 6
and 7 allow to be tested and observed in the future cold atom experiments.

\section{Conclusion}

We have studied the ground-state structures and spin textures of dipolar
BECs with DDI and Rashba SOC in a toroidal trap. We focus on the combined
effects of DDI and SOC on the ground-state properties of the system. A
ground-state phase diagram was obtained with respect to the SOC and DDI
strengths. It is shown that the DDI and the SOC, as two important
parameters, play a key role in determining the ground-state structure of the
system. Therefore they can be used to achieve the desired ground-state
phases and to control the phase transition between different ground states.
In addition, the system exhibits rich topological structures including
Anderson-Toulouse nucleated vortex state, curved vortex string composed of
vortices and antivortices, and necklace-like state with a giant vortex (or
antivortex) plus hidden antivortex (or vortex) necklace as well as further
ghost vortex (or antivortex) necklaces. Furthermore, the system sustains
exotic spin textures, such as different skyrmions, meron
(half-skyrmion)-antimeron (half-antiskyrmion) necklace, and complex meron
(half--skyrmion)-antimeron (half-antiskyrmion) lattice composed of a curved
meron (half-skyrmion) chain and an antimeron (half-antiskyrmion) pair. As
the ground states are stable against perturbation and have longer lifetime
in contrast to the other stationary states of the system, we expect that the
topological structures and spin textures of the system can be observed and
tested in the future experiments. These findings have greatly enriched our
new understanding of topological defects and spin defects in ultracold
atomic gases.

\begin{acknowledgment}

L.W. thanks W.Vincent Liu, Hui Zhai,Yongping Zhang and Malcolm
Jardine for useful discussions. This work was supported by the
National Natural Science Foundation of China (Grant Nos. 11475144 and 11047033),
the Natural Science Foundation of Hebei Province of China (Grant Nos. A2019203049 and A2015203037),
and Research Foundation of Yanshan University (Grant No. B846).

\end{acknowledgment}

\end{document}